\documentclass{kapproc} % Computer Modern font calls
\usepackage{psfig}
\kluwerbib
\begin{document}

\articletitle[ Dynamic stability of compact stars] {\centerline {Dynamic
stability of} \\ \centerline{ compact stars}}

\author{G.S.Bisnovatyi-Kogan
}
%% affil, email, and abstract are optional
\affil{Space Research Institute RAN, Moscow, Russia \\and\\Joint
Institute of Nuclear Researches, Dubna, Russia \footnote{Partial
funding provided by RFBR grant 02-02-16900.}}
\email{gkogan@mx.iki.rssi.ru}

%% optional, to supply a shorter version of the title for the running head:
%%\chaptitlerunninghead{}

\begin{abstract}
After some historical remarks we discuss different criteria of
dynamical stability of stars, and properties of the critical states
where dynamical stability is lost, leading to collapse with
formation of the neutron star or a black hole. At the end some
observational and theoretical problems related to quark stars are discussed.
\end{abstract}


\section{Introduction}

It was first noted by S Chandrasekhar in 1931 that the mass of the compact star at
very large densities, when electrons become relativistically degenerate cannot exceed
a value, which is now called as Chandrasekhar limit for white dwarfs (Chandrasekhar, 1931).
After discovery
of the neutron in 1932 the idea of the neutron star existence was developed. The matter
in these stars is so dense, that neutrons become relativistically degenerate,
again leading to the maximum mass of the neutron star. There are two important differences
between white dwarfs and neutron stars, influencing the stability and mass limit.
Neutron stars are about 6 orders of magnitude denser, than white dwarfs, making necessary
to use general relativity (GR) in description of gravitation, instead of newtonian gravity.
In GR the gravitation force is growing faster than $1/r^2$, what reduce a stability and
decrease the mass limit. On the other hand neutrons interaction become important at
densities exceeding the nuclear one. The repulsion forces between nucleons at large densities
lead to additional stabilization against collapse, increasing the limiting mass of the
neutron star. The influence of the second factor occurs to be more important, so at
present the limiting mass of the neutron star is taken about two times larger than that of the
white dwarf.

  Discovery  of the quark structure of the matter led to suggestion of possible existence
of quark stars, even more compact than neutron ones. In presence of indefiniteness concerning
quark structure of the matter it is not possible now to make definite statements
about possibility of existence or nonexistence of stable quark stars,
observational and theoretical
investigations on this topic are still in progress.

In this review I first make a historical excursus into the problem, mentioning the
results of the key works. Several criteria of stability are discussed, with the main
attention to the static criteria, and energetic method, which permit to obtain the
conclusion about the stability (sometimes approximately) in a most simple way.
Critical states of compact stars at the boundary of the dynamic stability are
considered, at which the star is becoming instable in the process of energy losses, and
collapse begins leading to formation of the neutron star or a  black hole. Physical
processes leading to a loss of stability are discussed. At the end some observations
and theoretical problems connected with quark stars are considered.

\section[Early development of the theory of compact stars: 1931-1965]
{Early development of the theory of compact stars: 1931-1965}

From the theory of polytropic gravitating gas spheres with the equation of state
$P=K\rho^{\gamma}$ it was known (Emden, 1907), that at $\gamma=4/3$ the equilibrium mass
has a unique value

\begin{equation}
\label{eq1}
M=4 \left(\frac{K\pi}{G}\right)^{3/2}M_3,
\end{equation}
where $M_3=2.01824$ is a non-dimensional mass, corresponding to the polytropic index
$n=3$, $\gamma=1+\frac{1}{n}$. Such a star is in a neutral equilibrium at any density.
Chandrasekhar (1931) first noticed, that at large
densities the electron degeneracy is becoming ultrarelativistic with the equation of state

\begin{equation}
\label{eq2}
P=\frac{\pi}{4} \left(\frac{3}{\pi}\right)^{1/3}\, \hbar c n_{\rm e}^{4/3},
\quad n_{\rm e}=\frac{\rho}{\mu m_p}.
\end{equation}
For some unknown reason Chandrasekhar (1931) took "$\mu$ equals the molecular weight, 2.5,
for a fully ionized material", corresponding to $K=3.619 \cdot 10^{14}$ and had obtained the
limiting mass from (\ref{eq1}) equal to $1.822\cdot 10^{33}$g$=0.91\, M_{\odot}$.
Landau (1932) noticed, that for most stable nuclei: He$^4$, C$^{12}$, O$^{16}$ etc.
the value of $\mu$, which is the number of nucleons on one electron, is equal to 2, and he obtained
the realistic value of the limiting mass $M_{lim}=1.4\, M_{\odot}$.

Soon after the neutron discovery (Chadwick, 24 Feb. 1932, letter to N.Bohr), "Landau
improvised the concept of neutron star" in discussion with Bohr (Rosenfeld, 1974).
The modern conception of the neutron star origin was due to Baade and Zwicky (1934),
which appeared as a result of their studies of supernovae explosions. Their conclusion:
"With all reserve we advance the view that supernovae represent the transitions from
ordinary stars into {\it neutron stars}, which in their final stages consist of extremely
closely packed neutrons", is an example of classic astrophysical predictions, finally proved
only 35 years later after discovery of the pulsar in the center of the Crab nebula
(Lovelace et al., 1968; Comella et al., 1969). Investigation of
properties of  matter at nuclear and larger densities
expected in neutron stars had been started in papers of Hund (1936),
Landau (1938) and Gamow (1938).

The first neutron star model in GR was constructed by Oppenheimer and Volkoff (1939).
They have derived equation of spherically symmetric stellar equilibrium in GR
using metric in the form

\begin{equation}
\label{eq3}
ds^2=-e^{\nu}\,dt^2+e^{\lambda}\,dr^2+r^2(d\theta^2+\sin^2\theta\,d\varphi^2)
\end{equation}
with equilibrium equations

\begin{equation}
\label{eq4}
  {dP\over dr}=-{G(\rho+P/c^2)(m+4\pi r^3P/c^2)\over r^2(1-2Gm/rc^2)},
\end{equation}
$$  {dm\over dr}=4\pi\rho r^2,
\quad  \rho=\rho_0\left(1+{E\over c^2}\right), \,\,\,
\rho(R)=0, \,\,\,M=m(R).$$
An ideal degenerate neutron gas at zero temperature was considered
with energy $E_n$, pressure $P_n$, and rest mass density $\rho_0$ as

$$
    E_n=\frac{m_n^4 c^5}{24 \pi^2 \hbar^3 \rho} g(y)={6.860\cdot 10^{35}\over \rho} g(y),
$$
\begin{equation}
\label{eq5}
 P_n=\frac{m_n^4 c^5}{24 \pi^2 \hbar^3 \rho_0} f(y)={6.860\cdot 10^{35}\over \rho_0} f(y),
\end{equation}
$$ \rho_0=\frac{m_n^4 c^3}{3\pi^2\hbar^3} y^3=
6.105\cdot 10^{15} y^3
$$
where

\begin{equation}
\label{eq6}
f(y)=y(2y^2-3)\sqrt{y^2+1}+3\sinh^{-1} y,\quad
\end{equation}
$$
g(y)=3y(2y^2+1)\sqrt{y^2+1}-3\sinh^{-1} y,\quad
y=\frac{p_{n,Fe}}{m_n c}.
$$
They had obtained the following parameters of the neutron stars of different
masses:

\begin{table}[ht]
\caption{Total mass of the neutron star model versus its radius,
calculated by Oppenheimer and Volkoff (1939).}
\begin{center}
\begin{tabular}{||c||c||l}
\hline
  %% On the next line is an example of how to get extra vertical space in
  %% a line: Use a \vrule with width 0pt and the height or depth that you
  %% want.
Total mass in Solar units\vrule height 14pt width 0pt depth 4pt
&Radius in km &\cr
\hline
\hline
0.30&21.1\vrule height 12pt width0pt&\cr
%%
%% On the next line, see how to line up numbers aligned on their decimal point
0.60&13.3&\cr
0.71&9.5&\cr
0.64&6.8&\cr
0.34&3.1&\cr
\hline
\end{tabular}
\end{center}
\end{table}
Between white dwarfs and neutron star there is a neutronization and transition from
normal matter with electrons and nuclei to superdense matter consisting of neutrons and
other strongly interacting particles: mesons and adrons. The continuous curve $M(\rho_c)$
in the whole region from white dwarfs to neutron stars was first constructed by Wheeler
(1958), where only neutrons have been considered in the superdense phase, like in
Oppenheimer and Volkoff (1939). The results of Wheeler (1958) are given in Fig.1.
The first neutron star model with a realistic equation of state, including mesons,
hyperons and nuclear interaction, was constructed by Cameron (1959). The following equilibrium
reactions between strongly interacting particles had been taken into account

$$n+n \rightarrow n+n+2\pi^0 \rightarrow \Lambda^0+\Lambda^0-2(177) \,\,{\rm MeV},
$$ $$
p+p \rightarrow p+p+2\pi^0 \rightarrow \Sigma^++\Sigma^+-2(251) \,\,{\rm MeV},
$$ $$
\Lambda^0+\Lambda^0 \rightarrow \Lambda^0+\Lambda^0 +\pi^0 \rightarrow
         \Sigma^0+\Sigma^0-2(77) \,\,{\rm MeV},
$$
\begin{equation}
\label{eq7}
n+n \rightarrow n+p+\pi^- \rightarrow p+\Sigma^- -256 \,\,{\rm MeV},
\end{equation}
$$n+\Lambda^0 \rightarrow p+\pi^- +\Lambda^0 \rightarrow p+\Xi^- -204 \,\,{\rm MeV},
$$ $$
\Lambda^0+\Lambda^0 \rightarrow \Lambda^0+\Lambda^0 +2\pi^0 \rightarrow
\Xi^0 +\Xi^0 - 2(195) \,\,{\rm MeV}.
$$
The nuclear interaction was considered according to Skryme (1959), see Fig.2. This
equation of state is not perfect, because at large densities it violates the casuality
principle, according to which the sound speed cannot exceed the light speed
(see Zeldovich, 1961). Neutron star models obtained by Cameron (1959) are
represented in Fig.3. The most important result of this work was the indication that
at realistic equation of state the limiting mass of the neutron star may exceed
the Chandrasekhar mass limit for white dwarfs. Therefore after a loss of stability
a stellar core with mass exceeding the Chandrasekhar mass limit may stop its contraction
forming a stable neutron star with enormous energy output, according to
Baade and Zwicky (1934). In the model of Oppenheimer and Volkoff (1939) the collapse
would not stop, because of the low mass limit of the neutron star.

Cameron (1959) had obtained that models at high densities had mo\-no\-to\-no\-us\-ly decreasing mass
asymptotically tending to a constant value. That was probably the result of a crudeness
of calculations, because the behavior of the curve $M(\rho_c)$ at large densities has
oscillating, and not monotonous dependence. Such oscillations had been present in calculations
of Ambartsumyan and Saakyan (1961), who also improved physical description of superdense
matter (Ambartsumyan and Saakyan, 1960). The nature of high density oscillations of mass in
GR equilibrium was explained by Dmitriyev and Kholin (1963) (see also
Misner and Zapolsky, 1964). They had shown that all models
beyond the first neutron star mass maximum on the curve $M(\rho_c)$ are unstable, and after each new
extremum the model acquires one new unstable mode. Later this problem was analyzed at length in the
book of Harrison et al. (1965). Detailed calculations of neutron star models with a realistic
equation of state had been done by Saakyan and Vartanyan (1964).

\begin{figure}[ht]
%\centerline{\epsfig{figure=cam1.eps,width=12cm}}
%\epsfsize=08cm {\centerline{\epsfbox{cam1.eps}}}
\centerline{
\psfig{file=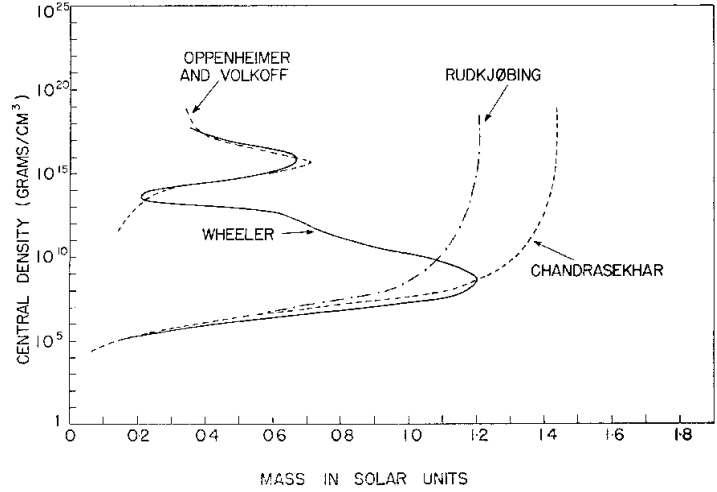,width=10cm,angle=-0}
}
\caption{The relation between central density and the mass of various degenerate star models.
Chandrasekhar's curve is for white dwarfs with a mean molecular weight 2 of atomic mass units.
Rudkj\'obing's curve is the same except for inclusion of the relativistic spin-orbit effects
Rudkj\'obing (1952).
The curve labeled "Oppenheimer and Volkoff" is for a set of neutron star models. The solid
line marked "Wheeler" is a set of models computed with a generalized equation of state,
from Cameron (1959)}.
\label{fig1}
\end{figure}

\begin{figure}[ht]
\centerline{
\psfig{file=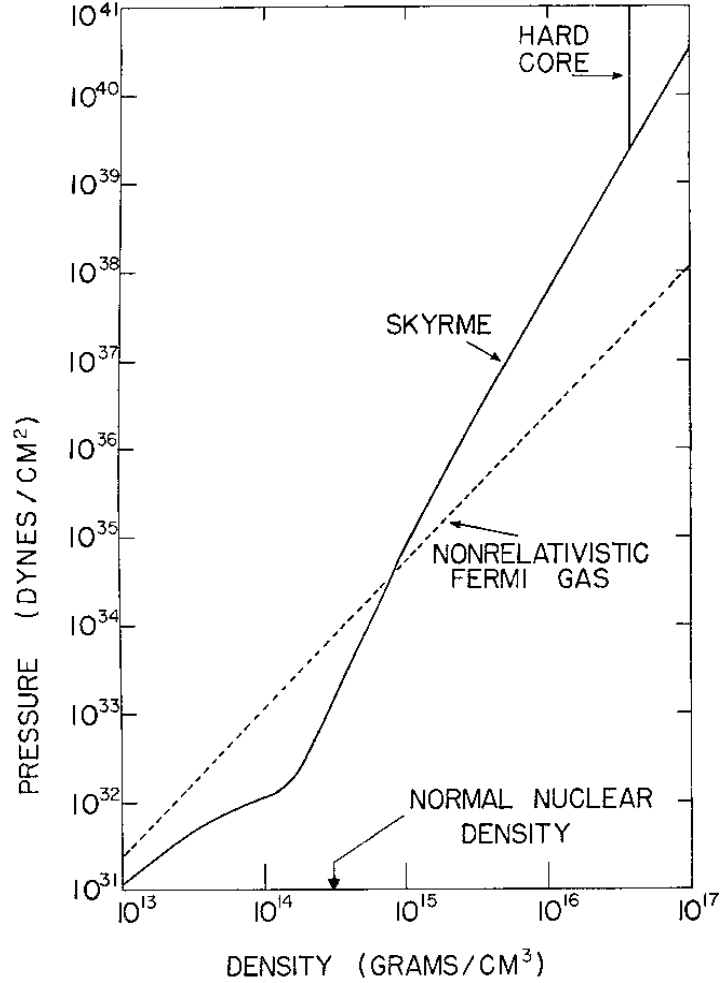,width=10cm,angle=-0}
}
\caption{Skryme's equation of state. The hard core modification is shown at the
upper right, and the dashed line is the equation of state for a non-relativistic
Fermi gas of neutrons, from Cameron (1959).}
\label{fig2}
\end{figure}

\begin{figure}[ht]
\centerline{
\psfig{file=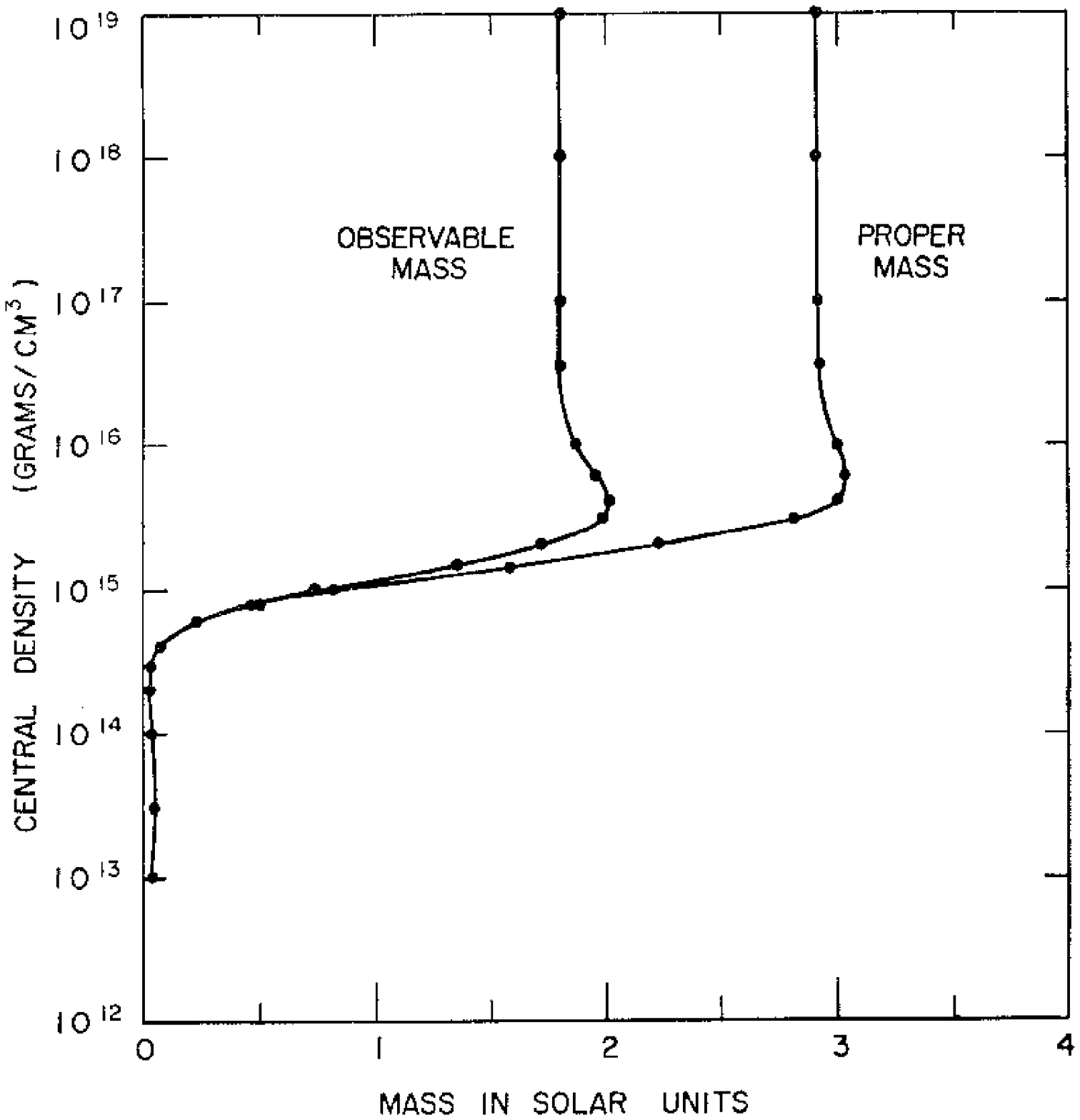,width=10cm,angle=-0}
}
\caption{The observable and proper masses of neutron star models with
non-ideal matter, from Cameron (1959).}
\label{fig3}
\end{figure}

\section{Criteria of hydrodynamic stability}

The exact approach to the problem of dynamic (linear) stability
is based on the solution of the equations for small perturbations,
and finding  eigenvalues and eigenfunctions of these equations.
In the conservative system the variation principle may be derived, which
determines the exact value of eigenfrequency if the exact eigenfunction is
known. In practice even use of an approximate linear eigenfunction often define
the eigenfrequency with a good precision. The variation principle for a spherically
symmetric star in GR was first derived by Chandrasekhar (1964) from the
equations of small perturbations, and later by Harrison et al. (1965) by variation
of the potential energy of the star.
  The exact method for determination of the dynamic stability of the star to the
adiabatic perturbations based on the sequence of static solutions was formulated by
Zeldovich (1963). He had shown, that in the extremum of the function $M(\rho_c)$, or
$M(R)$ along the sequence with a constant entropy $S$ adds or subtract one unstable mode.
On the curve of cold stars on Fig.1 the stability is lost in the first maximum due to
neutronization, or GR effects, but it is restored in the minimum, where stable
neutron star appears. The static criterium of stability was generalized for the case
of arbitrary rotating stars with arbitrary distribution of entropy over the star
in the paper of Bisnovatyi-Kogan and Blinnikov (1974). It was noted, that in presence
of three quantities conserved during adiabatic oscillations, the critical points are the
extrema of three functions $M(\rho_c)|_{s,j}$, $s(\rho_c)|_{M,j}$, $J(\rho_c)|{M,s}$.
Here along the sequences not only total values of stellar angular momentum $J$ and
entropy $S$ should be conserved, but also their distributions over the lagrangian
coordinate $ N$, which is the number of baryons inside the
cylindrical radius, $j(N)$ and $s( N)$. Note that in the newtonian case
lagrangian coordinate is coincides with lagrangian mass $m=\rho N=\rho_0 N$, but
differs from it in GR case. For polytropic equation of state
and angular momentum distribution similar to rigid rotation

\begin{equation}
\label{eq8}
P=K\rho^{1+1/n}, \,\,\, K={\rm const}\cdot e^s,\,\,\, j(m)=\frac{5}{2}\frac{J}{M}
[1-(1-m)^{2/3}]
\end{equation}
examples of application of the static criteria are represented in Fig. 4 from
Bisnovatyi-Kogan and Blinnikov (1974).

\begin{figure}
\centerline{\psfig{figure=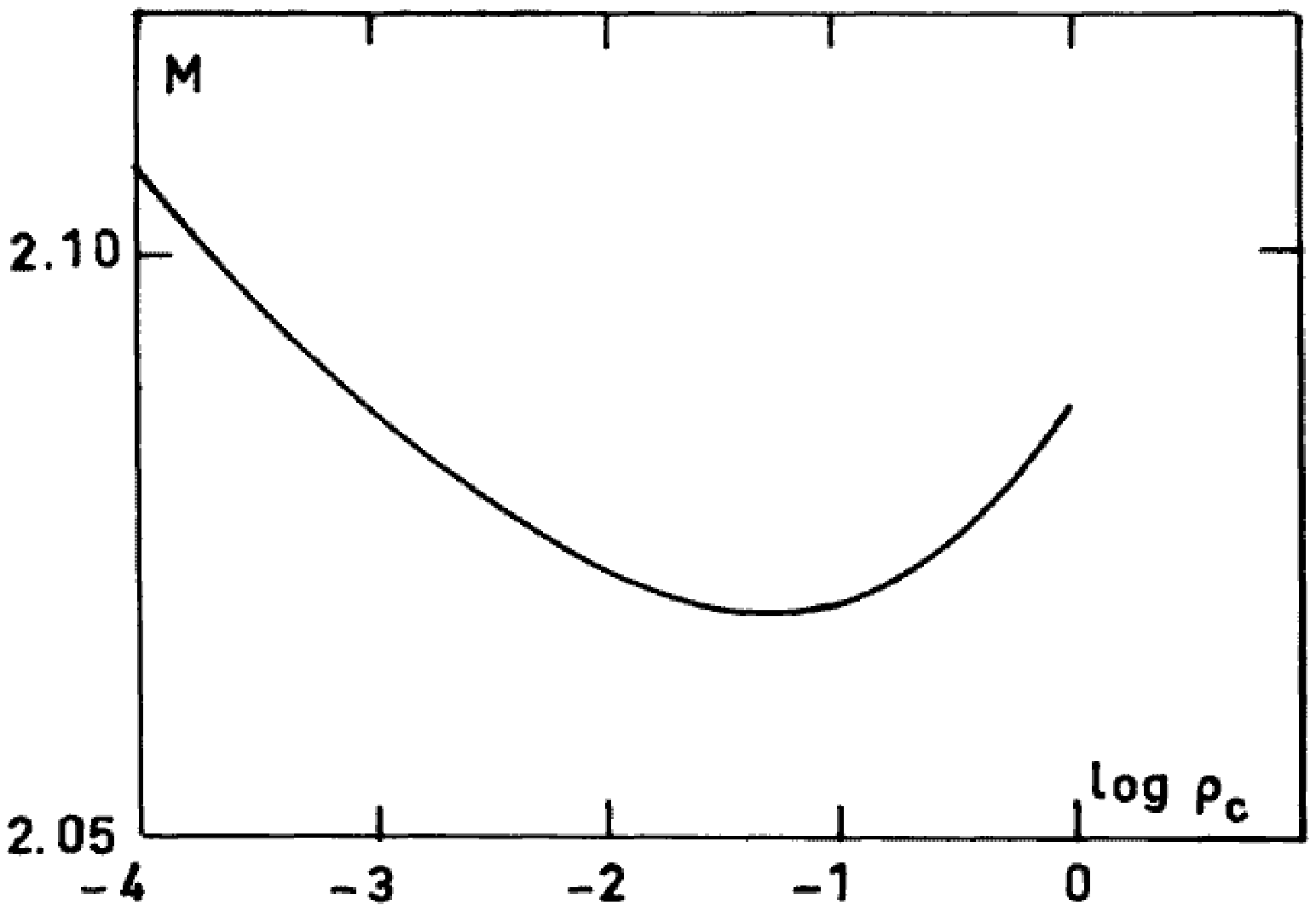,width=6.cm}\,
            \psfig{figure=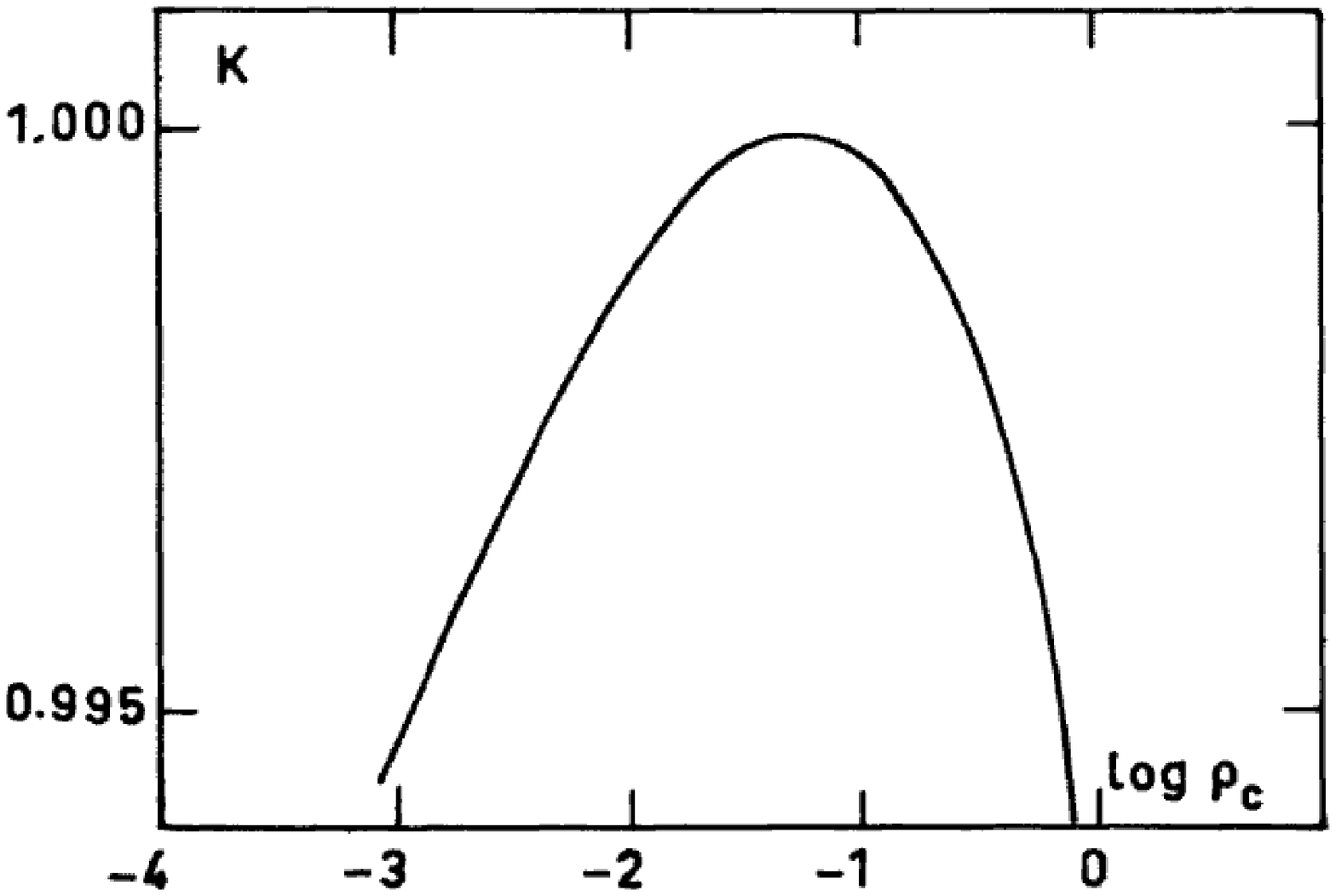,width=6.cm}
           }
\caption{The dependence of the mass $M$ on the central density for fixed
entropy and angular momentum, for isentropic models having a polytropic index
$n=3.03$. The angular momentum distribution is given in (\ref{eq8})
({\bf left}). The dependence of $K$ on central density for fixed
angular momentum distribution, corresponding to the minimum of mass in the left
figure, for isentropic models with polytropic index $n=3.03$ ({\bf
right}). From Bisnovatyi-Kogan and Blinnikov (1974). }
\label{fig4}
\end{figure}
Let us stress, that along the curve appropriate
for a static criterium of stability the rotation does not remain rigid, but only angular
momentum of each lagrangian loop should be conserved. It means that if you have a curve
of rigidly rotating stars with a fixed total angular momentum you should construct an
adjusted curve from each point of this curve with a constant angular momentum of a
lagrangian loop, which maximum indicated a loss of stability. The rigidly rotating
model is on the border of stability, when coincides with the maximum of the adjusted curve.
For a cold white dwarf with the equation of state

$$  \rho=Ax^3\left(2+a_1x+a_2x^2+a_3x^3\right),
$$
$$  P=B\left[x\left(2x^2-3\right)\left(x^2+1\right)^{1/2}
  +3\ln\left(x+\sqrt{1+x^2}\right)\right],
$$
\begin{equation}
\label{eq9}
  A=9.82\times 10^5~{\rm g~cm^{-3}},\quad
  B=6.01\times 10^{22}~{\rm erg~cm^{-3}},
\end{equation}
 $$ a_1=1.255\times 10^{-2},\quad
  a_2=1.755\times 10^{-5},\quad
  a_3=1.376\times 10^{-6},
$$
the adiabatic index is close to $n=3$ at the critical point, and the eigenfunction is
close the the linear one. At homologous transformation the rigid rotation is preserved,
so the adjusted curve for white dwarfs differs only slightly from the curve
of rigidly rotating models, see Fig.5 (left). For a more complicated equation of state
with highly variable adiabatic index, like in the equation of state

\begin{equation}
\label{eq10}
\rho(H)={1\over 2}\,\rho_1\left[\left(H\over H_1\right)^{0.3}
  +\left(H\over H_1\right)^{10}\right],
\end{equation}
$$  P(H)={1\over 2}\,\rho_1H_1\left[{(H/H_1)^{1.3}\over 1.3}
  +{(H/H_1)^{11}\over 11}\right].
$$
the value $\gamma_1=4{1 \over 3}$ at
$\rho=\rho_1$ and falls off steeply when $\rho>\rho_1$. The results of
calculations for a model with solid-body rotation and three associated
series of models are given in Fig.5 (right). The loss of stability occurs at the
point of intersection of the associated curve $D$ with the solid-body one,
this point coinciding with the maximum on the curve $D$. It can be seen
from Fig.5 (right) that the point of stability loss differs from the maximum on
the solid-body curve by almost 5\% with respect to $\rho_c$. An extension
of static criteria to general relativity and toroidal magnetic fields
is made in the paper of Bisnovatyi-Kogan and Blinnikov (1974).
Static criteria remains valid also in the presence of phase transitions.
Variational principle and other methods of investigation of stability in presence
of phase transitions are considered by Bisnovatyi-Kogan, Blinnikov and Shnol (1975).
The review of stellar stability methods is given in the book
of Bisnovatyi-Kogan (2002).

\section{Energetic method}

In the case of a complicated equation of state and entropy distribution along the star,
the static criteria is hard to apply. In this case it is more convenient to use
an approximate energetic method, which follows from the exact variation principle
where linear trial function is used for estimation of the eigenfrequency.

\begin{figure}
\centerline{\psfig{figure=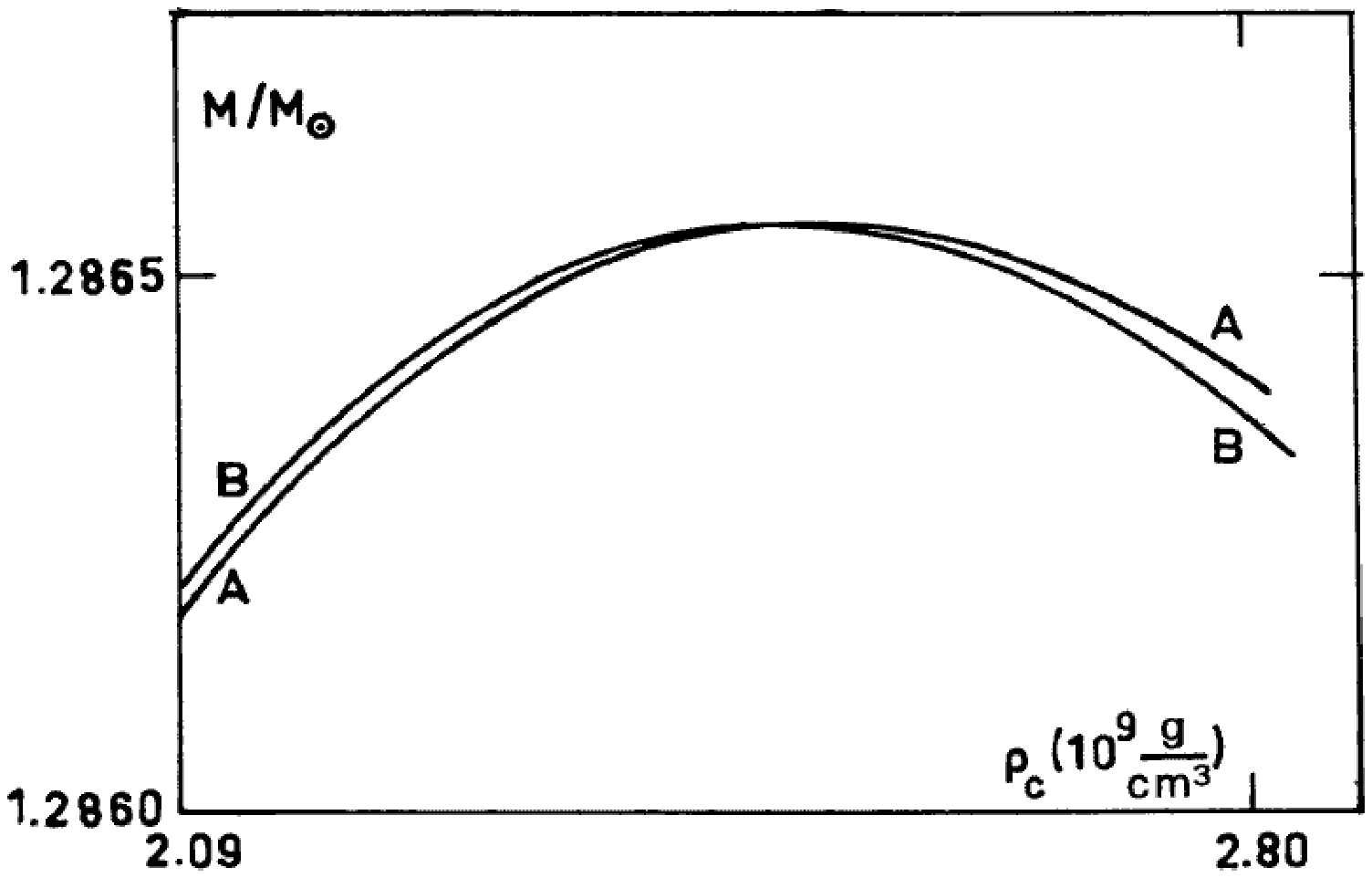,width=6.cm}\,
            \psfig{figure=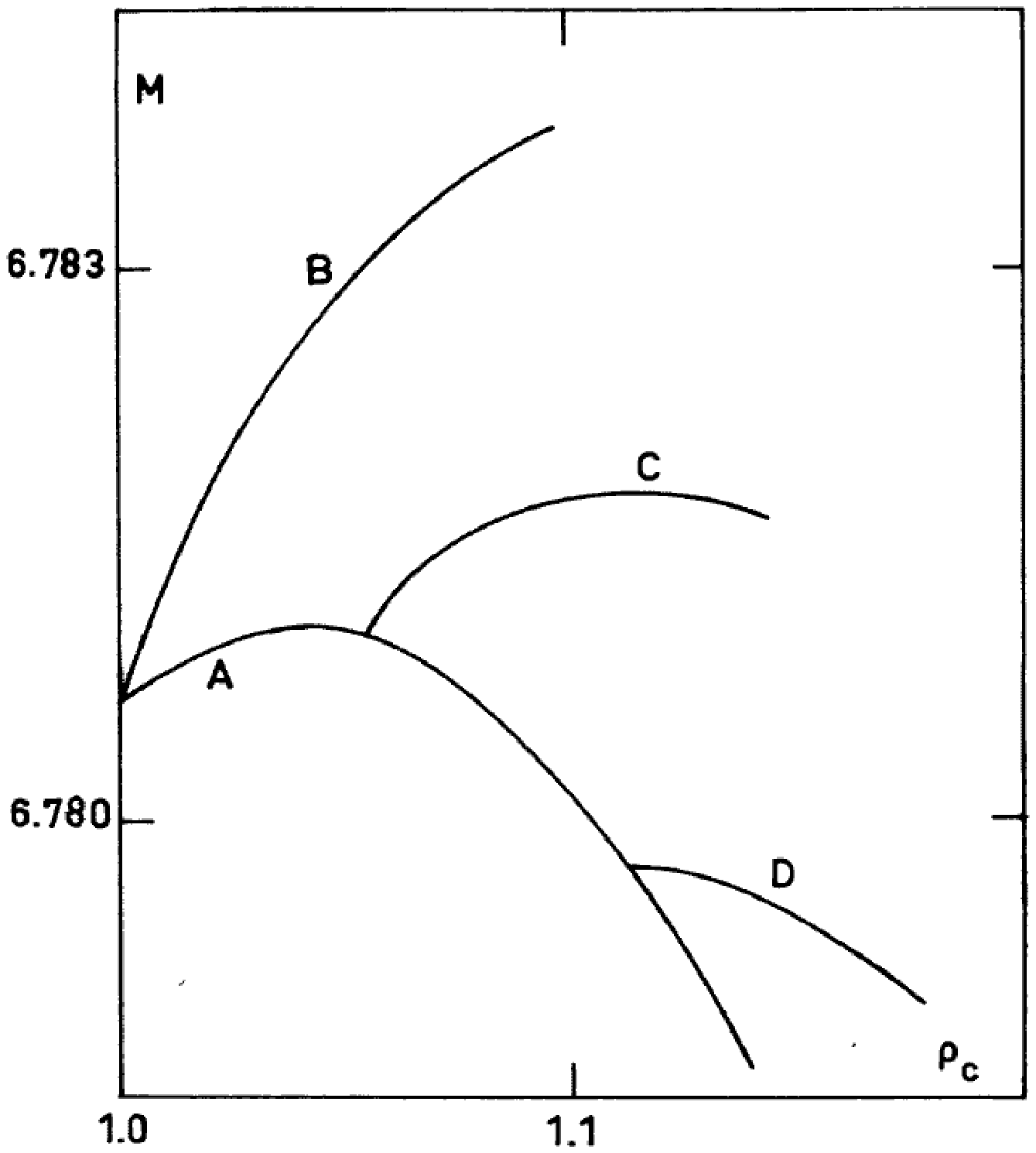,width=6.cm}
           }
\caption{ The dependence of the mass $M$ on the central density for
rigidly (Curve A) and differentially (Curve B) rotating white dwarfs
with the equation of state from (\ref{eq9}). The angular momentum distribution
of the Curve B is that of the extremal model of Curve A
({\bf left}). The dependence of the mass $M$ on the central density for
rigidly (Curve A) and differentially (Curves B,C and D) rotating models
with the equation of state from (\ref{eq10}). The angular momentum distribution
of the Curves B,C and D are those of the rigidly rotating models of the Curve A
at the points of its intersection with the Curves B,C and D. The curve C is
intersecting the Curve A in its maximum. The maximum of the Curve D coincides
with the point of intersection, which is a critical point for rigidly rotating
models.({\bf
right}). From Bisnovatyi-Kogan and Blinnikov (1974). }
\label{fig5}
\end{figure}
Energetic method was first applied for investigation of stability of supermassive
isentropic stars by Zeldovich and Novikov (1965) where first GR corrections to
the energy of such star have been found. This method was generalized for arbitrary
stars by Bisnovatyi-Kogan (1966). Equations have been derived, giving
approximate description of equilibrium and determining the critical point where
stability is lost. The first equation represents equilibrium condition, following from
first variation of the total potential energy of that star (gravitational plus internal
for nonrotating stars) equal to zero. The model with adiabatic index $n=3$, $\gamma=4/3$
is in a neutral stability with a linear eigenfunction. In application of the energetic method
the density distribution of the model is prescribed by the polytropic $n=3$ distribution,
$\rho=\rho_c \varphi(m/M)$, where $m$ is lagrangian mass coordinate, and $\varphi(m/M)$ is connected
with the Emden function corresponding to $n=3$, with a linear trial function $\delta r = \alpha r$,
$\alpha$=const.
The equation of state is arbitrary with the
prescribed entropy distribution over the lagrangian coordinate, and first GR correction to the
energy, including all terms $\sim R/R_g$, $R$ is stellar radius, $R_g=2GM/c^2$ is the
gravitational radius, are taken into account. The equilibrium equation in the approximate
energetic method is:

\begin{equation}
\label{eq11}
 3\rho_c^{4/3}\int_0^M P {dm\over \varphi\left(m\over M\right)}\,-0.639\,GM^{5/3}
-1.84\,{G^2M^{7/3}\over c^2}\,{\rho_c^{1/3}}=0.
 \end{equation}
Under the above mentioned conditions the second variation of the energy is
also reduced to the integral relation. Zero value of this relation
approximately corresponds to the
critical state of the loss of stability. We have

\begin{equation}
\label{eq12}
9\rho_c^{-5/3}\int_0^M
  \left(\gamma-{4\over 3}\right)P{dm\over \varphi(m/M)}
-1.84\,{G^2M^{7/3}\over c^2}=0.
 \end{equation}
These equations had been obtained by Bisnovatyi-Kogan (1966), using the expression
of the energy with the prescribed distributions of the density (Emden polytrope $n=3$)
and entropy (arbitrary) over the lagrangian coordinate $m/M$,

\begin{equation}
\label{eq13}
\epsilon=\int_0^M E(\rho,T)\,dm-\int_0^M{Gm\,dm\over r}
 -5.06\,{G^2M^3\over R^2c^2}
\end{equation}
$$=\int_0^M E(\rho,T)\,dm-0.639\,GM^{5/3}\,\rho_c^{1/3}
-0.918\,{GM^{7/3}\over c^2}\,\rho_c^{2/3},
$$ $$
dm=4\pi\rho r^2dr
$$
by differentiation over the central density. Critical parameters of iron isentropic stellar
cores for masses between 5 and 1000 $M_{\odot}$ at the point of the loss of stability
had been calculated by Bisno\-va\-t\-yi-Kogan
and Kazhdan (1966). The dependence of the critical central density on mass is represented
in Fig.6 (Bisnovatyi-Kogan, 2002) for a wide region of masses. At low masses, high densities
the stability in the iron core is lost due to neutronization. For larger masses the temperature
in the critical point is increasing, and stability is lost due to iron disintegration,
decreasing the adiabatic power, and making it $\gamma \le 1$. At masses around 1000 $M_{\odot}$
with decreasing density and temperature, the  adiabatic power becomes less than 4/3
mainly due to pair creation, and at farther increasing of mass the effects of GR are the
main reason of the loss of stability (Zeldovich and Novikov, 1965). At large masses the
dependence $\rho_c(M)$ is described by the relation

\begin{equation}
\label{eq14}
\rho_{c,{\rm cr}}(M)=2.4\times 10^{17}{1\over \mu^3}{\left(
  M_{\odot}\over M \right)}^{7/2}
  ~{\rm g~cm^{-3}},\quad
\end{equation}
where $\mu$  is the molecular weight.

Models of hot isentropic neutron stars had been calculated by Bis\-no\-va\-t\-yi-Kogan (1968), where
equilibrium between iron, protons and neutrons was calculated, and the ratio of
protons and neutrons was taken in the approximation of zero chemical potential of
neutrino. The stability was checked using a variational principle in full GR
(Chandrasekhar, 1964) with a linear trial function. The results of calculations,
showing the stability region of hot "neutron" stars are given in Fig.7.
Such stars may be called "neutron" only by convention, because they consist mainly from
nucleons with almost equal presence of neutrons and protons. The maximum of the
mass is about 70$M_{\odot}$, but from comparison of the total energies of hot neutron stars
with presupernova cores we may conclude, that only collapsing cores with masses less that
15$M_{\odot}$ have a chance to stop at the state of a hot neutron star, and larger masses
collapse directly to black holes.

\begin{figure}[ht]
\centerline{
\psfig{file=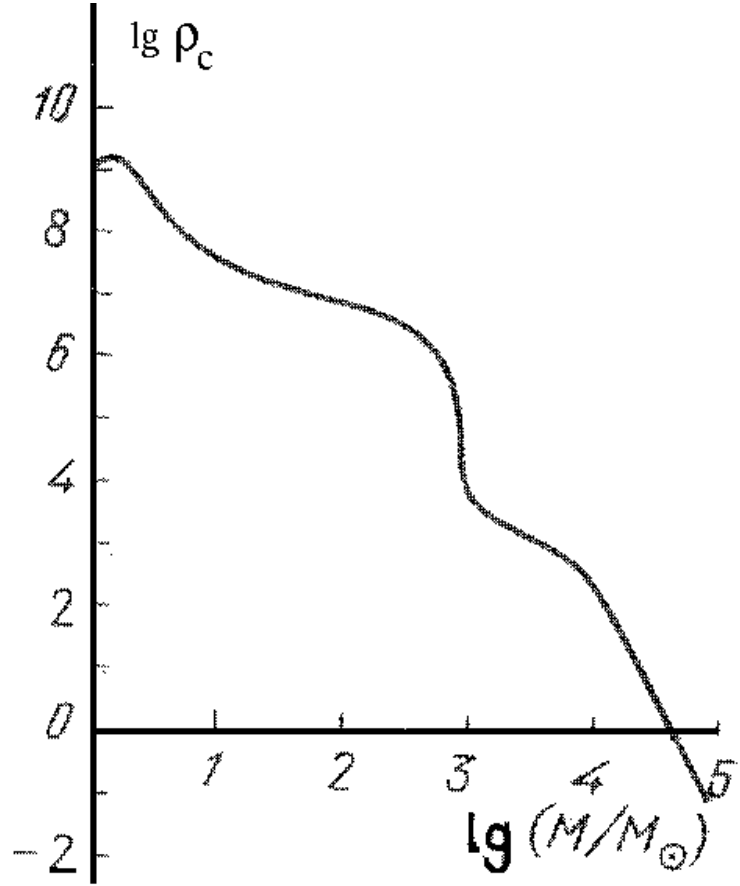,width=10cm,angle=-0}
}
\caption{Central density as a function of stellar core mass for the critical
states, from Bisnovatyi-Kogan (2002).}
\label{fig6}
\end{figure}

\begin{figure}[ht]
\centerline{
\psfig{file=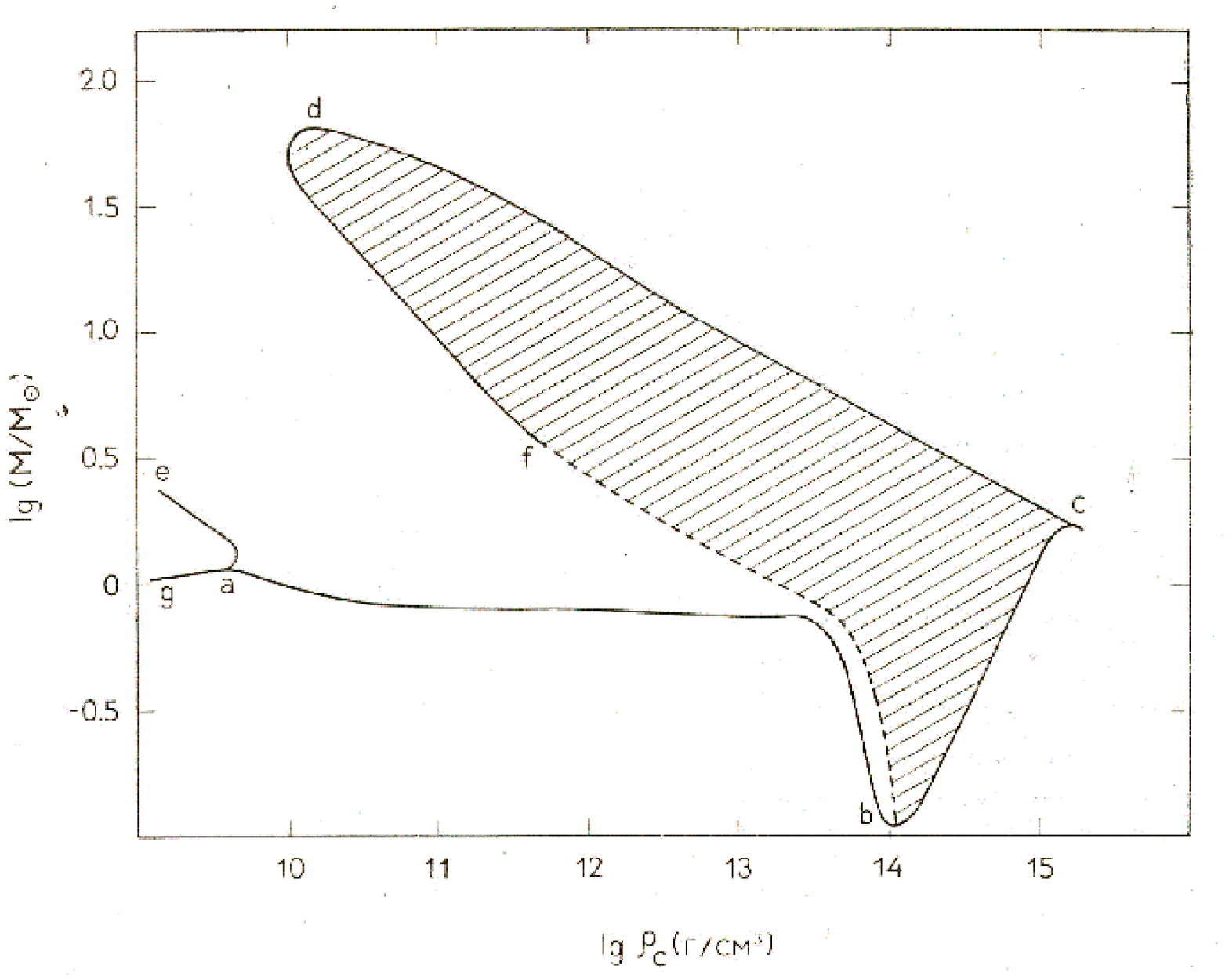,width=10cm,angle=-0}
}
\caption{Equilibrium stable configurations on the mass $M$, central
rest-mass density $\rho_c$ diagram for superdense isentropic stars (hatched).
The stars become stable at sufficiently large densities because of nucleon pressure,
and become unstable again at larger densities due to GR effects, from Bisnovatyi-Kogan (1968).}
\label{fig7}
\end{figure}

The energetic method was generalized for rotating stars in the paper of Bisnovatyi-Kogan,
Zeldovich and Novikov (1967), and in a more precise formulation by Bisnovatyi-Kogan and
Ruzmaikin (1973). The density distribution remains the same, and the rotational energy
term, written as a function of the total angular momentum was added. Rotational energy
of the whole star depends on the central density as $\sim R^{-2} \sim \rho^{2/3}$, like the
first GR correction term in (\ref{eq13}). Therefore next order GR corrections
had been taken into account
$\sim R^{-3}$, which include also GR corrections to the internal and rotational energies.
Calculations of the second GR corrections to the energy of nonrotating stars have been
first done by Vartanyan (1972). Energetic method with two GR corrections for rotating stars
was applied for investigation of stability of supermasive stars by Bisnovatyi-Kogan and
Ruzmaikin (1973). It was obtained, that for large angular momenta rotation stabilizes the
star against a collapse and prevent formation of a black hole. It is in correspondence with
a fact, that Kerr solution for a rotating black hole exist only for sufficiently small angular
momenta, and there are no black holes at larger values. Only radiation with $\gamma=4/3$ was
taken into account in the equation of state, what lead to absence of equilibrium configurations
at low angular momenta $j$. The curves for equilibrium configurations at different $j$ are
represented in Fig.8 on the plane: equilibrium entropy $s_{eq}$ - $x=R_g/R$.
The results are expressed in non-dimensional variables and
are valid for an arbitrary mass star with the adiabatic equation of state $P=K\,\rho^{4/3}$.
The energetic method gives asymptotically exact results for slowly rotating supermassive
stars, but even at large $j$ the results remain correct qualitatively.\footnote{Because an
approximate trial function does not give an absolute minimum of the energy functional, the
loss of stability happens, strictly speaking, "before" the point obtained by the
energetic method (EM), when going along the sequence of decreasing entropy or angular
momentum. It means, that the model is definitely unstable in the EM critical point.}

\begin{figure}[ht]
\centerline{
\psfig{file=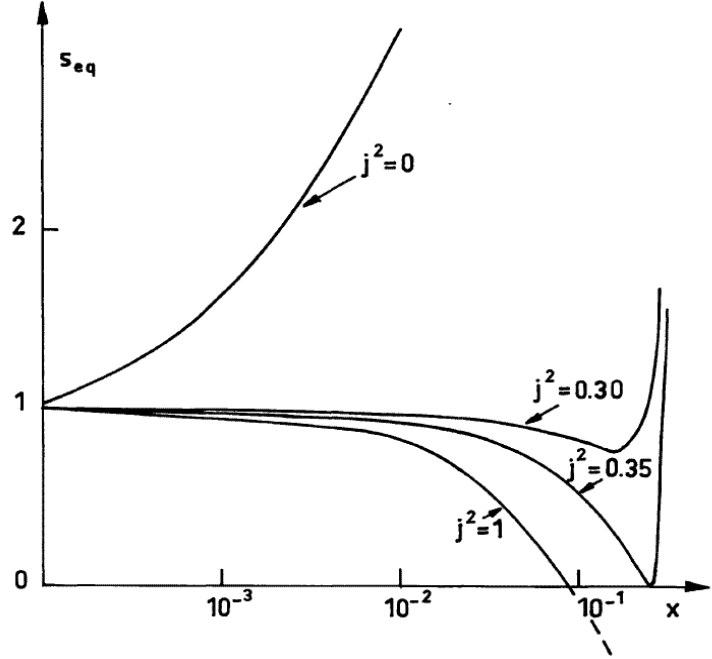,width=10cm,angle=-0}
}
\caption{The curves $s_{eq}$ which characterizes the equilibrium entropy
versus of $x=\frac{R_g}{R}$ for different values of the angular momentum $j$. The broken
part corresponds to the nonphysical states $s_{eq}<0$. The parts of curves to the left of
minima correspond to the stable states,
from Bisnovatyi-Kogan and Ruzmaikin (1973).}
\label{fig8}
\end{figure}

\section{Search of quark stars}

At very high densities the equilibrium lowest energy state corresponds to quarks.
However because of big indefiniteness of the theory the condition at which the
quark state becomes preferable lays in a big density interval. Moreover, the process
of transition between nucleon and quark state is also not definite, and even at very
large densities the time of this transition may be enormously long.
First the idea of a possibility of quark stars was suggested by
Ivanenko and Kurdgelaidze (1965). Depending on the
theory parameters there is a possibility of 3 types of quark stars:
strange stars (SS), which consists of strange matter which may be stable at nuclear density
with a zero pressure; hybrid stars with hadronic envelope,
which has a core of pure quarks (QC), or
has a mixed core of hadrons and quarks (MC). There are different approaches to describe a
quark matter, see Aguirre and De Paoli (2002), Andersen and Strickland (2002),
Burgio et al. (2002), Sedrakian and Blaschke (2002),
Kohri et al. (2003), Lugones and Horvath (2003),
Alford (2003), Mishustin et al. (2003).
The results of calculation of hadronic (H) and
quark stellar models (SS, QC and MC) in Hard-Dense-Loop approach are represented in Fig.9 from
Thoma et al. (2003), where one of a model parameters is changing.
The free quarks exist in the state
of deconfined quarks, and the density when deconfined quarks become energetically preferable
is also rather indefinite (Berezhiani et al., 2003).

\begin{figure}[ht]
\centerline{
\psfig{file=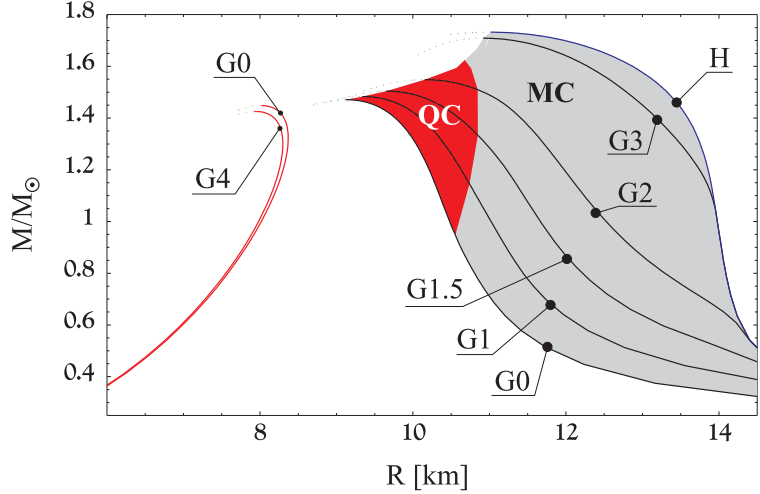,width=10cm,angle=-0}
}
\caption{Mass-radius relation for pure strange quark matter stars (left) and
hybrid stars (right). G0 - G4 models of hybrid stars corresponding to different
parameters of the model. H: pure hadron star, QC: star has a quark core, MC: star
has a mixed core, from Thoma et al. (2003).}
\label{fig9}
\end{figure}

Larger scattering of the properties of quark stars is obtained by
Andersen and Strickland (2002), who used the same Hard-Dense-Loop approach, but with
wider variation of parameters. Their models of quark stars are represented in Fig.10.

\begin{figure}[ht]
\centerline{\psfig{file=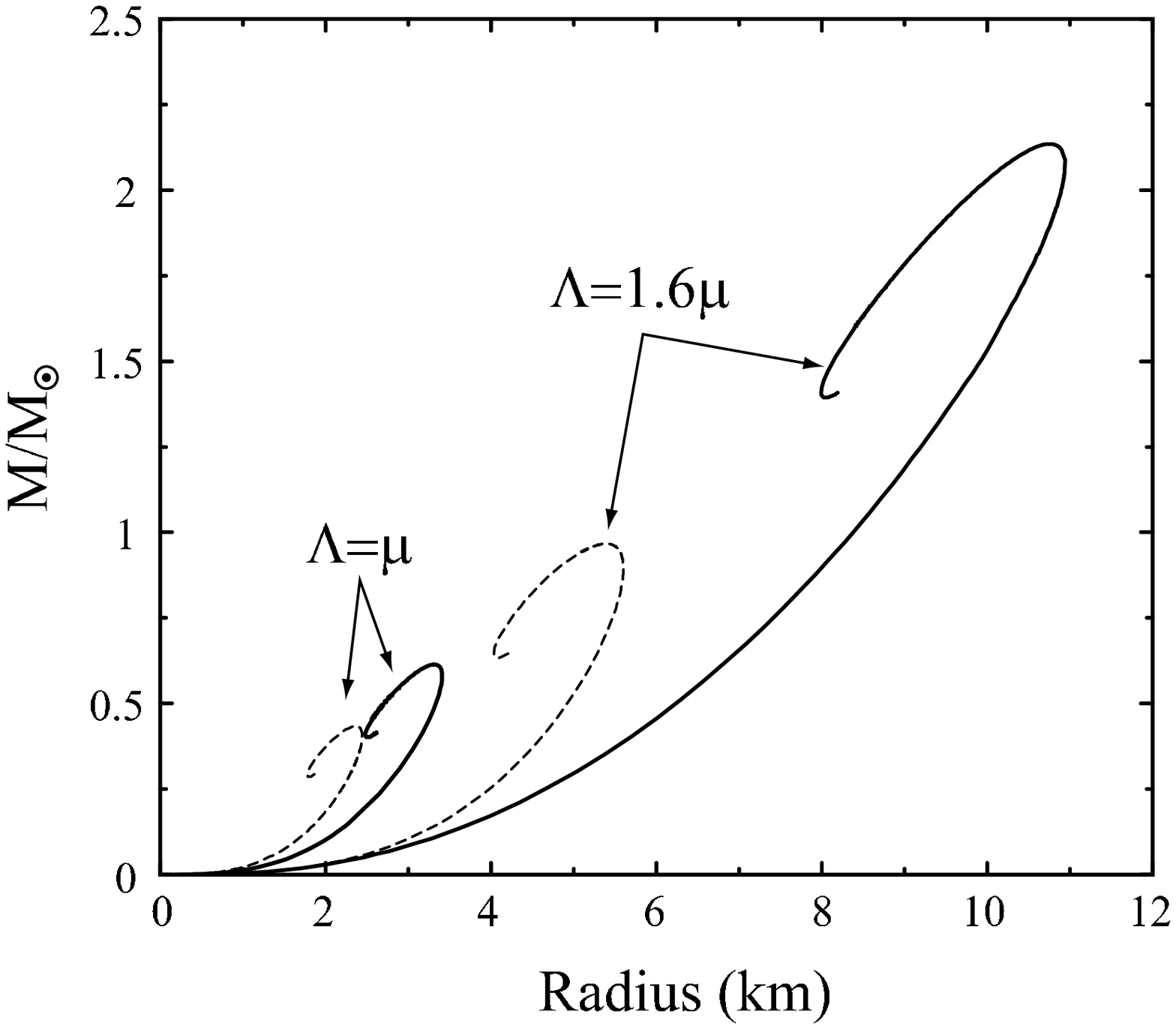,width=10cm}}
\caption{
Mass-radius relation for a quark star with $\Lambda/\mu=1.6$ and $\Lambda/\mu=1$.
The weak-coupling results for the same choice of renormalization scales are shown as
dashed lines, from Andersen and Strickland (2002).
}
\label{fig10}
\end{figure}

In presence of such problems with theory, only observations could give an answer
about the possibility os existence of quark stars.  Even in presence of theoretical
indefiniteness, it is impossible to have a neutron (pure hadronic) star with a radius
smaller than about 10 kilometers. Observational discovery of a compact star with
considerably smaller
radius would be an important evidence that this  star is not pure hadronic.
Such claim was made by Drake et al. (2002) basing on deep Chandra observations
of the isolated compact star RX J1856.5-3754. They wrote: "We argue that the derived
interstellar medium neutral hydrogen column density of $8\times
10^{19}\leq N_H \leq 1.1\times 10^{20}$ cm$^{-2}$ favours the larger
distance from two recent HST parallax analyses, placing RX J1856.5-3754 at $\sim
140$~pc instead of $\sim 60$~pc. ... The combined
observational evidence---a lack
of spectral and temporal features, and an implied radius
$R_\infty=3.8$-8.2~km that is too
small for current neutron star models---points to a more
compact object, such as allowed for quark matter equations of state."
Subsequent analysis of
observational data did not confirm this conclusion. Thoma et al. (2003) came to
contrary result of an unusually large value of the radius even for a neutron star:
"Combining such a X-ray spectrum with the optical
spectrum, one finds a black-body radius $R_\infty >17$ km, indicating a stiff
equation of state, which would exclude a strange quark matter star or
even hybrid star. This is a rather conservative lower limit for the radius
since a black-body emitter is the most efficient radiator." Similar conclusion was
done by Turolla et al. (2003):
"While a quark star may
be a conceivable option, present
observations do not necessarily demand this
solution and more conventional scenarios involving a neutron star
are certainly possible. Neutron star models based on a
two-temperature surface distribution can account for both the
X-ray and optical emission, giving at the same time
acceptable values for the stellar radius."
Meanwhile a new candidate for a quark (strange) star in the X-ray source
4U 1728-34 was suggested by Bombaci (2003), and farther discussion about
this object is expected.

So far no reliable candidate for a quark (strange) star was found by observations.
The theoretical predictions remain uncertain, and could explain either existence or
nonexistence of quark stars. The solution of intriguing problem of cosmological
gamma ray burst (GRB) origin was proposed by Berezhiani et al (2003) basing on
energy production during transformation of the neutron star into a stable quark (strange)
star. This model could explain the connection of GRB with supernovae explosions with a
formation of a neutron star, and subsequent huge energy release during transition to a state of
a quark star producing GRB. An attractive feature of this model is a possibility of
explanation of any time delay between SN and GRB explosions, which depends on the
transition time from hadron to quark state, and is very sensitive to parameters.
In this model the time delay may be arbitrary long, so majority of SN explosions
in accordance with observations do
not produce GRB, because this time exceeds the cosmological Hubble time.

\section{Summary}
1. Formation of quark (strange) stars does not follow unambiguously from the theory,
which may be compatible either with existence or nonexistence of these objects.

2. There are no firm observational contradictions to the conventional model of the hadronic
neutron star.

3. Speculations about connection of the cosmological GRB with transition from the hadronic to
quark star seems to be interesting, because they explain connection between GRB and supernovae
explosion, with arbitrary time delay between these events, including very large,
exceeding the Hubble time.

\begin{acknowledgments}
The author wish to thank D. Sedrakyan, D. Blaschke, G. Hadjan, Yu. Vartanyan,
E. Chubaryan and A. Sadoyan
for support and hospitality during the conference; and I.D. Novikov for useful
discussion.
\end{acknowledgments}

\end{document}